\documentclass[11pt]{amsart}
\usepackage{latexsym,amssymb,amsmath,amscd,amsthm}
\numberwithin{equation}{section}
\theoremstyle{remark}

\newcommand{\bq}{\begin{equation}}
\newcommand{\bea}{\begin{array}}
\newcommand{\eea}{\end{array}}

\newcommand{\ga}{\alpha}

\newcommand{\gD}{\Delta}
\newcommand{\gl}{\lambda}
\newcommand{\gL}{\Lambda}
\newcommand{\gb}{\beta}

\newcommand{\mf}{\mathfrak}
\newcommand{\mc}{\mathcal}

\newcommand{\go}{\omega}

\newcommand{\gag}{\gamma}
\newcommand{\gd}{\delta}
\newcommand{\pp}{\partial}

\newcommand{\tl}{\tilde}
\newcommand{\na}{\nabla}

\newcommand{\bs}{\blacksquare}

\newcommand{\bgs}{\bigstar}

\newcommand{{\DDD}}{D\!\!\!\!\!\!-}

\title{REMARKS ON FISHER INFORMATION}

\author{Robert Carroll\\University of Illinois, Urbana, IL 61801}

\date{July, 2008\thanks{email: rcarroll@math.uiuc.edu}}

\begin{document}

\bibliographystyle{plain}

\begin{abstract} 
Some situations are discussed where subquantum oscillations in momentum arise in 
connection with Fisher information and the quantum potential.
\end{abstract}

\maketitle
\tableofcontents


\section{INTRODUCTION}
\renewcommand{\theequation}{1.\arabic{equation}}
\setcounter{equation}{0}

In \cite{c7} we derived a formula for Fisher information (FI) based on the thermalization of
Gr\"ossing \cite{gros} for the quantum potential.  We want to embellish this here in various ways
and indicate connections to other work.  For Fisher information we refer to \cite{c3,f1,f2,f4} and for
related work on thermodynamics and diffusion see e.g. \cite{c3,crow,ebsv,garb,gabr,glts,ohpz}.

\section{BACKGROUND}
\renewcommand{\theequation}{2.\arabic{equation}}
\setcounter{equation}{0}

We consider first a standard Schr\"odinger equation (SE) in the form $({\bf 2A})\,\,i\hbar\psi_t=
-(\hbar^2/2m)\psi''+V\psi$ where $'\sim\pp_x\,\,or\,\,\na_x$ so that for $\psi=Rexp(iS/\hbar)$
and $P=R^2=|\psi|^2$ one has
\bq\label{2.1}
({\bf A})\,\,S_t+\frac{1}{2m}(S')^2+V-\frac{\hbar^2}{2m}\frac{R''}{R}=0;\,\,
({\bf B})\,\,P_t+\frac{1}{m}(PS')'=0
\end{equation}
For 1-D situations the quantum potential (QP) is written in the form $({\bf 2B})\,\,Q=-(\hbar^2/2m)(R''/R)=-(\hbar^2/2m)
(\pp_x^2\sqrt{\rho}/\sqrt{\rho})$ for $\rho=mP$.  One often sets $p=(1/m)S'$ (momentum)
and evidently $\int R^2dx=\int Pdx=1$.  It is seen directly that
\bq\label{2.2}
\frac{\pp^2\sqrt{\rho}}{\sqrt{\rho}}=\frac{1}{4}\left[\frac{2\rho''}{\rho}-\left(\frac{\rho'}{\rho}\right)^2\right]
\end{equation}
(where $\rho$ could be replaced by $P$) and hence with $P=0$ for $|x|\geq x_0$
\bq\label{2.3}
\int P\frac{\pp^2_x\sqrt{P}}{\sqrt{P}}dx=\frac{1}{4}\int\left(2P''-\frac{(P')^2}{P}\right)dx=
-\frac{1}{4}\int\frac{(P')^2}{P}dx=-\frac{1}{4}FI
\end{equation}
(FI means Fisher information).  Hence in particular, writing Q in terms of P,
\bq\label{2.4}
\int PQdx=\frac{\hbar^2}{8m}\int\frac{(P')^2}{P}dx=\frac{\hbar^2}{8m}FI
\end{equation}
Write now for differential entropy $({\bf 2C})\,\,{\mf S}=-\int \rho log(\rho)dx$ and note that
\bq\label{2.5}
{\mf S}=-\int mP log(mP)dx=-m\int P[log(m)+log(P)]dx=
\end{equation}
$$=-m\int Plog(P)dx-mlog(m)$$
since $\int Pdx=1$.  Consequently
\bq\label{2.6}
\pp_t{\mf S}=-m\int [P_tlog(P)+P_t]dx=-m\int P_t(log(P)+1)dx
\end{equation}
From (2.1) we have $P_t=-(1/m))PS')'$ so
\bq\label{2.7}
\pp_t{\mf S}=\int(PS')'[log(P)+1]dx=-\int (PS')\frac{P'}{P}dx=-\int S'P'dx
\end{equation}
\\[3mm]\indent
{\bf REMARK 2.1.}
At this point one often refers to diffusion or hydrodynamical processes and thinks of $S'=p=m\dot{q}$
where $v=\dot{q}$ represents some collective velocity field related to a free Brownian motion for example
with say $v=-(\hbar/2m)(\na P/P)$ involving an osmotic velocity (cf. \cite{c3,garb,gabr}).  
Then (2.1B) becomes $({\bf 2D})\,\,
P_t+(Pv)'=0$ and it is interesting to note that this form of velocity involving $\na P/P$ or 
$\na\rho/\rho$ comes up in the form of momentum fluctuations or perturbations in work of
Hall-Reginatto \cite{hall,hkr,hlal,regi} on the exact uncertainty principle and in Crowell \cite{crow}
in a fluctuation context.  Indeed in \cite{crow} one takes $({\bf 2E})\,\,\gd\rho=(\hbar/2i)(\na\rho/\rho)$ with
\bq\label{2.8}
<\gd\rho>=\int \rho\gd\rho dx=\frac{\hbar}{2i}\int \na\rho dx=0;\,\,<(\gd \rho)^2>=c\int\left[
\frac{(\na\rho)^2}{\rho}\right]dx\sim FI
\end{equation}
Thus Q can be written as
\bq\label{2.9}
Q=-\frac{\hbar^2}{4m}\left[\frac{1}{2}(\gd p)^2-\frac{\gD\rho}{\rho}\right]=
\frac{\hbar^2}{4m}\left[\na(\gd p)+\frac{1}{2}(\gd p)^2\right]
\end{equation}
The exact uncertainty principle involves the characterization of quantum fluctuations as being
generated by momentum fluctuations $\na P/P$ under certain quite general assumptions.  This
principle can be applied in many interesting circumstances (cf. \cite{c3,carr,hall,hkr,hlal,regi}).
$\hfill\bs$
\\[3mm]\indent
{\bf REMARK 2.2.}
We note also, going back to (2.7) with $P'\sim\na P$, when (as in certain diffusion processes) $S'=-(\hbar/2m)(P'/P)$, then 
$({\bf 2F})\,\,\pp_t{\mf S}=(\hbar/2m)\int[(P')^2/P]dx\sim \hat{c}FI$ showing that $\pp_t{\mf S}
\geq 0$ and exactly how it varies.  This feature also arises in studying Ricci flow and 
information on this is provided in \cite{c6,khlo,klbd,mulr,perl,topp}.
Let $P(x)$ be a probability distribution on {\bf R} or ${\bf R}^3$; we work in {\bf R} for
simplicity.  Consider a differential entropy ${\mf S}=-\int P\,log(P)dx$ and assume one wants
to specify some value $\bar{A}=\int A(x)P(x)dx$.  Then consider extremizing
\bq\label{2.10}
\tl{{\mf S}}=-\int Plog(P)dx+\gl(1-\int Pdx)+\ga(\bar{A}-\int APdx)
\end{equation}
Recall $log(1+x)\sim x$ for $x$ small and write
\bq\label{2.11}
\gd \tl{S}=-\int\left[(P+\gd P)log(P+\gd P)-Plog(P)\right]dx-\gl\int  \gd Pdx-\ga\int A\gd Pdx=
\end{equation}
$$=-\int\left\{(P+\gd P)log\left[P\left(1+\frac{\gd P}{P}\right)\right]-Plog(P)\right\}dx
-\int (\gl+\ga A)\gd Pdx=$$
$$\approx-\int \gd P(x)\left[log(P)+1+\gl+\ga A\right]dx$$
Since $\gd P$ is arbitrary one obtains $({\bf 2G})\,\,log(P)+1+\gl+\ga A=0$ or
$P=exp[-1-\gl-\ga A]=(1/Z)exp(-\ga A)$ where $Z=exp(1+\gl)$
and $\int Pdx=1\Rightarrow Z=\int exp[-\ga A(x)]dx$.  Here a suitable candidate for A would be
the total energy of a system and if S is determined by P alone with $P\sim p=(1/m)S'$
as above then A would have to be the fluctuation energy determined by the quantum potential.
$\hfill\bs$

\section{SUBQUANTUM THERMODYNAMICS}
\renewcommand{\theequation}{3.\arabic{equation}}
\setcounter{equation}{0}

We go here to \cite{gros} for a derivation of the SE from vacuum fluctuations and
diffusion waves in sub-quantum thermodynamics (cf. also \cite{bgzk,boyr,
gacn,garb,gabr,grsg,lpso,wsen} for background).  Gr\"ossing work in \cite{gros}
specifies the energy in quantum mechanics arising from sub-quantum fluctuations 
via nonequilibrium thermodynamics.
The ideas are motivated and discussed at great length in e.g. papers 1, 2, and 4 in \cite{gros}
and we only summarize here following \cite{gros}-6.
To each particle of nature is attributed an energy $E=\hbar\go$ for some kind of angular frequency
$\go$ and one can generally assume that ``particles" are actually dissipative systems maintained
in a nonequilibrium steady-state by a permanent source of kinetic energy, or heat flow, which
is not identical with the kinetic energy of the particle, but an additional contribution.  Thus it
is assumed here that $({\bf 3A})\,\,E_{tot}=\hbar\go+[(\gd p)^2/2m]$ where $\gd p$ is the
additional fluctuating momentum component of the particle of mass $m$.  Similarly the
particle's environment is considered to provide detection probability distributions which can be
modeled by wave-like intensity distributions $I(x,t)=R^2(x,t)$ with $R$ being the wave's
real valued amplitude; thus one assumes $({\bf 3B})\,\,P(x,t)=R^2(x,t)$ with $\int Pd^nx=1$
(note $x\sim {\bf x}$).
In \cite{gros}-1 it was proposed to merge some results of nonequilibrium thermodynamics
with classical wave mechanics in such a manner that the many microscopic degrees of freedom
associated with the hypothesized sub-quantum medium can be recast into the more 
``macroscopic" properties that characterize the wave-like behavior on the quantum level.
Thus one considers a particle as being surrounded by a large ``heat bath" so that the momentum
distribution in this region is given by the usual Maxwell-Boltzmann distribution.  This corresponds
to a ``thermostatic" regulation" of the reservoir's temperature which is equivalent to saying that
the energy lost to the thermostat can be regarded as heat.  This leads to emergence at the
equilibrium-type probability density ratio $({\bf 3C})\,\,[P(x,t)/P(x,0)]=exp[-(\gD{\mc Q})/kT)]$
where T is the reservoir temperature and $\gD{\mc Q}$ the exchanged heat between the particle
and its environment.  The conditions ({\bf 3A})-({\bf 3C}) are sufficient to derive the SE.
Thus first, via Boltzmann, the relation between heat and action is given via an action function 
$S=\int (E_{kin}-V)dt$ with $\gd S=\gd\int E_{kin}dt$ via
\bq\label{3.1}
\gD{\mc Q}=2\go\gd S=2\go[\gd(S)(t)-\gd S(0)]
\end{equation}
(cf. \cite{gros} for more details).
Next the kinetic energy of the thermostat is $kT/2$ per degree of freedom and the average
kinetic energy of an oscillator is $(1/2)\hbar\go$ so equality of average kinetic energies demands 
$({\bf 3D})\,\,kT/2=\hbar\go/2$ or $\hbar\go=kT=1/\gb$.
Combining ({\bf 3C}), (3.1), and ({\bf 3D}) yields then 
\bq\label{3.2}
P(x,t)=P(x,0)e^{-\frac{2}{\hbar}[\gd S(x,t)-\gd S(x,0)]}
\end{equation}
leading to 
a momentum fluctuation
\bq\label{3.3}
\gd p(x,t)=\na(\gd S(x,t))=-\frac{\hbar}{2}\frac{\na P(x,t)}{P(x,t)}
\end{equation}
and an additional kinetic energy term
\bq\label{3.4}
\gd E_{kin}=\frac{1}{2m}\na(\gd S)\cdot\na(\gd S)=\frac{1}{2m}\left(\frac{\hbar}{2}\frac{\na P}{P}\right)^2
\end{equation}
The action integral then becomes 
\bq\label{3.5}
A=\int L d^nxdt=\int P(x,t)\left[\pp_tS+\frac{1}{2m}\na S\cdot\na S+\frac{1}{2m}\left(
\frac{\hbar}{2}\frac{\na P}{P}\right)^2+V\right]
\end{equation}
We emphasize here that
$$(\bgs\bgs)\int P(\na S\cdot \gd p)d^nx=\int P (\na S\cdot\na(\gd S))d^nx=0$$
(i.e. the fluctuations terms $\gd p$ are uncorrelated with the momentum $p\sim\na S$).
Now one uses the Madelung form $({\bf 3E})\,\,\psi=Rexp[(i/\hbar)S]$ where $R=\sqrt{P}$ to
obtain $({\bf 3F})\,\,[\na\psi/\psi]^2=[\na P/2P]^2+[\na S/\hbar]^2$ leading to (3.5) in the form
\bq\label{3.6}
A=\int Ldt=\int d^nx dt\left[|\psi|^2\left(\pp_tS+V\right)+\frac{\hbar^2}{2m}|\na\psi|^2\right]
\end{equation}
(cf. \cite{v2} for $|\psi|^2\sim P$).  
Then via $|\psi|^2\pp_tS=-(i\hbar/2)(\psi^*\dot{\psi}-\dot{\psi^*}\psi)$ one has
$({\bf 3G})\,\,L=-(i\hbar/2)(\psi^*\dot{\psi}-\dot{\psi^*}\psi)+(\hbar^2/2m)
\na\psi\cdot\na\psi^*+V\psi\psi^*$ leading to the SE $({\bf 3H})\,\,i\hbar\pp_t\psi=[-(\hbar^2/2m)
\na^2+V]\psi$ along with the ``modified" Hamilton-Jacobi equation
\bq\label{3.7}
\pp_tS+\frac{1}{2m}(\na S)^2+V+Q=0;\,\,Q=-\frac{\hbar^2}{4m}\left[\frac{1}{2}\left(\frac{\na P}
{P}\right)^2-\frac{\gD P}{P}\right]=-\frac{\hbar^2}{2m}\frac{\gD R}{R}
\end{equation}
Then define $({\bf 3I})\,\,{\bf u}=(\gd {\bf p}/m)=-(\hbar/2m)(\na P/P)$ and $k_{{\bf u}}=-(1/2)
(\na P/P)=-(\na R/R)$ so that Q can be rewritten as
\bq\label{3.8}
Q=\frac{m{\bf u}\cdot {\bf u}}{2}-\frac{\hbar}{2}(\na\cdot{\bf u})=\frac{\hbar^2}{2m}(k_{{\bf u}}
\cdot k_{{\bf u}}-\na\cdot k_{{\bf u}})
\end{equation}
Using (3.1) and (3.2) one can also write $({\bf 3J})\,\,{\bf u}=(1/2\go m)\na{\mc Q}$.
\\[3mm]\indent
Generally a steady state oscillator in nonequilibrium thermodynamics corresponds to a kinetic
energy at the sub-quantum level providing the necessary energy to maintain a constant oscillation
frequency $\go$ and some excess kinetic energy resulting in a fluctuating momentum contribution
$\gd p$ to the momentum $p$ of the particle (note $p\sim{\bf p}$).  Similarly a steady state
resonator representing a ``particle" in a thermodynamic environment will not only receive kinetic
energy from it but in order to balance the stochastic influence of the buffeting momentum
fluctuations it will also dissipate heat into the environment.  There is a vacuum fluctuation theorem
(VFT) from \cite{gros}-1 which proposes that larger energy fluctuations of the 
oscillating system correspond to higher probability of heat dissipated into the environment
(rather that absorbed).  The corresponding balancing velocity is called (after Einstein) the
``osmotic" velocity.  Thus recalling the stochastic ``forward" movement ${\bf u}\sim (\gd{\bf p}/m)$,
the current ${\bf J}=P{\bf u}$ has to be balanced by $-{\bf u}$, i.e. ${\bf J}=-P{\bf u}$.  Putting
({\bf 3I}) into the definition of the ``forward" diffusive current ${\bf J}$ and recalling the
diffusivity $D=\hbar/2m$ one has $({\bf 3K})\,\,{\bf J}=P{\bf u}=-D\na P$ and when combined
with the continuity equation $\dot{P}=-\na\cdot{\bf J}$ this gives $({\bf 3L})\,\,\pp_tP=D\na^2P$.
Here ({\bf 3K}) and ({\bf 3L}) are the first and second of the Fick laws of diffusion and ${\bf J}$ is
called the diffusion current.
\\[3mm]\indent
Returning now to ({\bf 3J}) one defines $\gD{\mc Q}={\mc Q}(t)-{\mc Q}(0)<0$ and maintaining
heat flow as positive one writes $-\gD {\mc Q}$ for heat dissipation and puts this in ({\bf 3J})
to get the osmotic velocity $({\bf 3M})\,\,\bar{{\bf u}}=-{\bf u}=D(\na P/P)=-(1/2\go m)\na{\mc Q}$
with osmotic current $({\bf 3N})\,\,\bar{{\bf J}}=P\bar{{\bf u}}=D\na P=-(P/2\go m)\na{\mc Q}$.
As a corollary to Fick's second law one has then
\bq\label{3.9}
\pp_tP=-\na\cdot\bar{{\bf J}}=-D\na^2P=\frac{1}{2\go m}[\na P\cdot\na{\mc Q}+P\na^2{\mc Q}]
\end{equation}
Next one looks for a thermodynamic meaning for the quantum potential Q.  Take first $Q=0$
and look at the osmotic velocity ({\bf 3M}) that represents the heat dissipation from the particle
into its environment.  From (3.8) one has $({\bf 3O})\,\,(\hbar/2)(\na\cdot{\bf u})=(1/2)(m{\bf u}\cdot
{\bf u})$.  Inserting now ({\bf 3M}) instead of ({\bf 3J}) yields the thermodynamic corollary of
a vanishing QP as $({\bf 3P})\,\,\na^2{\mc Q}=-(1/2\hbar\go)(\na{\mc Q})^2$.  Returning to
(3.9) put first ({\bf 3D}) into ({\bf 3L}) to get $({\bf 3Q})\,\,P=P_0exp[-(\gD{\mc Q}/\hbar\go)]$
and one obtains then from (3.9)
\bq\label{3.10}
\pp_tP=\frac{P}{2\go m}\left[\na^2{\mc Q}-\frac{(\na{\mc Q})^2}{\hbar\go}\right]\Rightarrow
\pp_tP=-\frac{P}{2\go m}\na^2{\mc Q}
\end{equation}
(via ({\bf 3P})).  Now from ({\bf 3Q}) one has also $({\bf 3R})\,\,\pp_tP=-(P/\hbar\go)\pp_t{\mc Q}$
so comparison of (3.10) and ({\bf 3Q}) yields $({\bf 3S})\,\,\na^2{\mc Q}-(1/D)\pp_t{\mc Q}=0$.
This is nothing but a classical heat equation obtained by the requirement 
that the quantum potential $Q=0$; it shows that even for free particles both in the quantum and classical case one can identify a heat
dissipation process emanating from the particle.  A non-vanishing quantum potential then is a
means of describing the spatial and temporal dependencies of the corresponding thermal flow
in the case that the particle is not free.
\\[3mm]\indent
Various particular solutions to ({\bf 3S}) are indicated for the case $Q=0$ as well as when
$Q\ne 0$.  Examples are discussed with a view toward resolving a certain 
``particle in a box" problem of Einstein.  In \cite{gros}-1 one concentrated on the momentum
fluctuations $\gd p$ generated from the environment to the particle while in \cite{gros}-3 one
develops the idea of excess energy developed as heat from the particle to its environment
(which is described via the quantum potential).  In fact one can rewrite (3.7) as
\bq\label{3.11}
\pp_tS+\frac{1}{2m}(\na S)^2+V+\frac{\hbar^2}{4m}\left[\na^2\tl{{\mc Q}}
-\frac{1}{D}\pp_t\tl{{\mc Q}}\right]=0
\end{equation}
where $\tl{{\mc Q}}={\mc Q}/\hbar\go$.  This gives wonderful insight into the nature and role of
the quantum potential.
\\[3mm]\indent
Note that one has achieved a ``thermalization" of the quantum potential (QP)
in the form
\bq\label{3.12}
Q=\frac{\hbar^2}{4m}\left[\na^2\tl{{\mc Q}}-\frac{1}{D}\pp_t\tl{{\mc Q}}\right]
\end{equation}
where $\tl{{\mc Q}}={\mc Q}/\hbar\go=\ga {\mc Q}$ is an expression of heat and $D=\hbar/2m$ is a 
diffusion coefficient (note $\ga=\gb$ as in ({\bf 3D}). 
In \cite{c7} we show that, as a corollary,
one can produce a related thermalization of Fisher information (FI) which should have
interesting consequences.
Thus in ({\bf 3N}) one uses a formula
\bq\label{3.13}
\frac{\na P}{P}=-\frac{1}{2\go mD}\na{\mc Q}=-\frac{1}{\go\hbar}\na{\mc Q}=-\gb\na{\mc Q}
\end{equation}
and this leads one to think of 
$\na log(P)=-\ga\na{\mc Q}=-\na(\ga{\mc Q})$ with a possible solution
\bq\label{3.14}
log(P)=-\ga{\mc Q}+c(t)\Rightarrow P=exp[-\ga{\mc Q}+c(t)]=\hat{c}(t)e^{-\ga{\mc Q}}
\end{equation}
(note also $({\bf 3T})\,\,{\mc Q}(t)\sim 2\go\gd S(t)$ and $(2/\hbar)\gd S(t)\sim (1/kT){\mc Q}(t)=\gb{\mc Q}(t)$).
Now Fisher information (FI) is defined via ($dx\sim dx^3$ for example)
\bq\label{3.15}
F=\int\frac{(\na P)^2}{P}dx=\int P\left(\frac{\na P}{P}\right)^2dx
\end{equation}
and one can write Q as in (3.7).
Consequently (since $\int \gD Pdx=0$)
\bq\label{3.16}
\int PQdx=-\frac{\hbar^2}{8m}\int\frac{(\na P)^2}{P}dx=-\frac{\hbar^2}{8m}F
\end{equation}
Then as in \cite{c7} one can write formally, first using (3.12) and $\ga=1/\go\hbar$
\bq\label{3.17}
F=-\frac{8m}{\hbar^2}\int PQdx=-\frac{8m}{\hbar^2}\int P\frac{\hbar^2}{4m}\left[\na^2\tl{{\mc Q}}
-\frac{1}{D}\pp_t\tl{{\mc Q}}\right]dx=
\end{equation}
$$=-2\ga\int P\left[\na^2{\mc Q}-\frac{2m}{\hbar}\pp_t{\mc Q}\right]dx$$
and secondly, using (3.14) and (3.13) (recall also $\ga=\gb$)
\bq\label{3.18}
F=\int P\left(\frac{\na P}{P}\right)^2dx=\gb^2\hat{c}(t)\int e^{-\gb{\mc Q}}(\na{\mc Q})^2dx
\end{equation}
In view of the thermal aspects of gravity theories now prevalent it may perhaps
be suggested that connections of quantum mechanics to gravity may best be handled thermally.
There may also be connections here to the emergent quantum mechanics of \cite{elze,hoof}.

\section{REMARKS ON FLUCTUATIONS}
\renewcommand{\theequation}{4.\arabic{equation}}
\setcounter{equation}{0}

We go here to \cite{crow} which contains a rich lode of important material on quantum
fluctuations and some penentrating insight into physics (but some proofreading 
seems indicated).  We will try to rewrite
some of this in a more complete manner.  Thus consider a  SE for $\psi=Rexp(iS/\hbar)$
with momentum operator $\hat{p}$ so that
\bq\label{4.1}
\hat{p}\psi=p\psi=\left(\na S+\frac{\hbar}{i}\frac{\na R}{R}\right)\psi\Rightarrow
p\sim <p>+\gd p
\end{equation}
which identifies $(\hbar/i)(\na R/R)$ as a fluctuation $\gd p$.  Recall now that the SE
is $({\bf 4A})\,\,i\hbar\psi_t=-(\hbar^2/2m)\gD \psi+V\psi$ and setting $\rho=R^2=
\psi^*\psi$ one has
\bq\label{4.2}
\pp_t\rho=\frac{i\hbar}{2m}[\psi^*\gD\psi-(\gD\psi^*)\psi]=\frac{i\hbar}{2m}\na\cdot
(\psi^*\na\psi-\na(\psi^*)\psi)
\end{equation}
Now in polar form $\psi=Rexp(iS/\hbar),\,\,\rho=R^2$ this becomes (calculating in 1-D for
simplicity)
\bq\label{4.3}
(Re^{iS/\hbar})'=R'e+\frac{RiS'}{\hbar}e\sim \na(Re^{iS/\hbar})=\na Re^{iS/\hbar}+\frac
{iR\na S}{\hbar}e^{iS/\hbar}
\end{equation}
\bq\label{4.4}
\psi^*\na\psi-(\na\psi^*)\psi=R\na R+\frac{iR^2\na S}{\hbar}-R\na R+\frac{iR^2\na S}{\hbar}
=\frac{2i\rho\na S}{\hbar}\Rightarrow
\end{equation}
$$\Rightarrow \pp_t\rho=\frac{i\hbar}{2m}\na\left(\frac{2i\rho\na S}{\hbar}\right)=-\frac{1}{m}
\na(\rho\na S)$$
This agrees with (2.8) in \cite{crow}, but (2.7), (2.9), and (2.10) in \cite{crow} are somewhat ``strange".  Now
the quantum potential is 
\bq\label{4.5}
Q=-\frac{\hbar^2}{2m}\frac{\gD\rho^{1/2}}{\rho^{1/2}}=-\frac{\hbar^2}{4m}\left[\frac{1}{2}
\left(\frac{\na\rho}{\rho}\right)^2-\frac{\gD\rho}{\rho}\right]=
\frac{\hbar^2}{4m}\left[\frac{\gD\rho}{\rho}-\frac{1}{2}\left(\frac{\na\rho}{\rho}\right)^2\right]
\end{equation}
(cf. ({\bf 2B}) and $\rho=R^2$ so $\rho'=2RR'$ and $2(R'/R)=\rho'/\rho$; hence in (4.1) one has
$({\bf 4B})\,\,\gd p\sim(\hbar/2i)(\na\rho/\rho)$.  There is equivalent material about 
Fokker-Planck equations in \cite{c3} so we omit the discussion in \cite{crow}.  One notes
that $({\bf 4C})\,\,<\gd p>\sim\int \rho\gd p dx=(\hbar/2i)\int \na\rho dx=0$ whereas
$<(\gd p)^2>\sim c\int[(\na \rho)^2/\rho]dx\sim Fisher\,\,information$.  Here the quantum potential can also be written as
\bq\label{4.6}
Q=-\frac{\hbar^2}{4m}\left[\frac{1}{2}(\gd p)^2-\frac{\gD\rho}{\rho}\right]=
\frac{\hbar^2}{4m}\left[\na(\gd p)+\frac{1}{2}(\gd p)^2\right]
\end{equation}
since $(\rho'/\rho)'=(\rho''/\rho)-[(\rho')^2/\rho^2]\sim \gD \rho/\rho=\na(\gd p)+(\gd p)^2$.
Note also that the exact uncertainty principle of Hall-Reginatto (\cite{hall,hkr}), developed
at length in \cite{c3}, is based on momentum fluctuations $p=\na +\gd p$ with $<\gd p>=0$.
If one writes $V(q)= V(<q>)+\na_qV(<q>)\gd q$ as a function of position and recalls
that the quantum HJ equation has the form $({\bf 4D})\,\,S_t+(p^2/2m)+V+Q=0$
(cf. \cite{c3}) then this can be rewritten in terms of fluctuations as
\bq\label{4.7}
S_t+\frac{1}{2m}<p^2>+V(<q>)+\na_qV(<q>)\gd q+Q(p,\gd p);
\end{equation}
$$Q=
\frac{\hbar^2}{4m}\left(-i\hbar\na\cdot\gd p+\frac{1}{2}(\gd p)^2\right)$$

\section{EXAMPLES}
\renewcommand{\theequation}{5.\arabic{equation}}
\setcounter{equation}{0}

We go now to \cite{fpps} with $({\bf 5A})\,\,\int dx\,p(x)=1$ and $({\bf 5B})\,\,I[p]=
\int dx F_I(p)$ where $F_I(p)=p(x)[(p')/p]^2$.  Assume that there are known 
\bq\label{5.1}
<A_j>=\int dx A_j(x)p(x)\,\,\,(j=1,\cdots,M)
\end{equation}
Then one uses the principle of extreme physical information (EPI) to find the probability
distribution $p=p_I$ extremizing $I[p]$ subject to prior conditions $<A_j>$.  Jaynes used
the Shannon functional $F=-plog(p)$ with $({\bf 5C})\,\,S[p]=-\int dx plog(p)$
but here one uses the Fisher extremization with 
\bq\label{5.2}
\gd_p\left[I[p]-\ga<1>-\sum_1^M\gl_i<A_i>\right]=0\equiv 
\end{equation}
$$\equiv \gd_p\left[\int dx\left(F_I[p]-
\ga p-\sum_i^M\gl_iA_ip\right)\right]=0$$
Variation leads to
$$\gd\int\frac{p^{'2}}{p^2}dx\approx \int\left[-\frac{p^{'2}}{p^2}\gd p+\frac{2p'}{p}\gd p'
\right]dx
\sim \int \left[-\frac{p^{'2}}{p^2}-\pp\left(\frac{2p'}{p}\right)\right]\gd pdx$$
\bq\label{5.3}
\int dx\gd p\left[(p)^{-2}(p')^2+\pp_x\left(\frac{2}{p}p'\right)+\ga+\sum_1^M\gl_iA_i\right]
=0
\end{equation}
which implies, via the arbitrary nature of $\gd p$
\bq\label{5.4}
\left[(p)^{-2}(p')^2+\pp_x\left(\frac{2}{p}p'\right)+\ga+\sum_1^M\gl_iA_i\right]=0
\end{equation}
The normalization condition on $p$ makes $\ga$ a function of the $\gl_i$ and one
lets $p_I(x,\gl_i)$ be a solution of (5.4).  Then the extreme Fisher information is
$({\bf 5D})\,\,I=\int dx p^{-1}_I[p'_I]^2$.  Now one can simplify (5.4) via
$({\bf 5E})\,\,G(x)=+\ga+\sum_1^M\gl_iA_i(x)$ and write (5.4) as
\bq\label{5.5}
[\pp_xlog(p_I)]^2+2\frac{\pp^2log(p_I)}{\pp x^2}+G(x)=0
\end{equation}
Then introduce $p_I=\psi^2$ and $({\bf 5F})\,\,v(x)=\pp_xlog(\psi(x))$ so that (5.5) 
becomes $({\bf 5G})\,\,v'(x)=-[(1/4)G(x)+v^2(x)]$ which is a Riccati equation.  Setting
\bq\label{5.6}
u(x)=exp\left[\int^xdx\frac{d log(\psi)}{dx}
\right]=\psi
\end{equation}
makes (5.5) into a Schr\"odinger like equation 
\bq\label{5.7}
-\frac{1}{2}\psi''(x)-\frac{1}{8}\sum \gl_iA_i(x)\psi(x)=\frac{\ga}{8}\psi
\end{equation}
where $({\bf 5H})\,\,U(x)=(1/8)\sum_1^M\gl_iA_i(x)$ is an effective potential
(cf. \cite{f5,frsf,fpps,frdn,nifr}).  Note that $\psi$ is defined here completely via $p_I$ so any
quantum motion is automatically generated by fluctuation energy (i.e. $S\sim\gd S$).
\\[3mm]\indent
Consider now a situation with one function $A_i\,\,\,({\bf 5I})\,\,\bar{A}=\int pAdx$. 
Then from (5.2)-(5.5)
\bq\label{5.8}
p^{-2}(p')^2+\pp_x\left(\frac{2p'}{p}\right)+\ga+\gl A=0
\end{equation}
Now, following \cite{fpps}, one translates the Legendre structure of thermodynamics
into a Fisher context.  Thus from ({\bf 5D}), integrating by parts implies 
\bq\label{5.9}
\frac{\pp I}{\pp\gl}=\int dx\frac{\pp p_I}{\pp\gl}\left[-p_I^{-2}\left(\frac{\pp p_I}{\pp x}\right)^2
-\frac{\pp}{\pp x}\left(\frac{2}{p_I}\frac{\pp p_I}{\pp x}\right)\right]
\end{equation}
where $p_I$ is a solution of (5.8).  Comparing (5.8) to (5.9) one has
\bq\label{5.10}
\frac{\pp I}{\pp\gl}=\int dx\frac{\pp p_I}{\pp\gl}\left[\ga+\gl A\right]
\end{equation}
Then on account of normalization ($\int p_Idx=1$) 
\bq\label{5.11}
\frac{\pp I}{\pp \gl}=\gl\frac{\pp}{\pp\gl}\int dx\, p_IA(x)\equiv \frac{\pp I}{\pp \gl}=\gl
\frac{\pp}{\pp\gl}<A>
\end{equation}
which is a generalized Fisher-Euler theorem.  Here the term $\int dx\ga\pp_{\gl} p_I\sim\ga\pp_{\gl}\int dx p_I=0$ via $\int p_Idx=1$.
The thermodynamic counterpart of (5.11)
is the derivative of the entropy with respect to mean values.  Thus $I=I(\gl),\,\,
p_I=p_I(\gl)$, and via normalization $\ga=\ga(\gl)$.  Thus $\gl$ and $<A>$ play 
reciprocal roles within thermodynamics and one introduces a generalized thermodynamic
potential as a Legendre transform of I, namely
\bq\label{5.12}
\gL=I(<A>)-\gl<A>
\end{equation}
Then, using (5.11)
\bq\label{5.13}
\frac{\pp\gL}{\pp \gl}=\frac{\pp I}{\pp<A>}\frac{\pp<A>}{\pp\gl}-\gl\frac{\pp<A>}{\pp\gl}
-<A>=-<A>
\end{equation}
and one has a summary collection of formulas
\bq\label{5.14}
\gL=I-\gl<A>;\,\,\frac{\pp \gL}{\pp\gl}=-<A>;\,\,\frac{\pp I}{\pp<A>}=\gl;
\end{equation}
$$\frac{\pp\gl}{\pp<A>}=\frac{\pp^2 I}{\pp<A>^2};\,\,\frac{\pp<A>}{\pp\gl}=-\frac
{\pp^2\gL}{\pp\gl^2}$$
and we recall (5.11) in the form $({\bf 5J})\,\,
\frac{\pp I}{\pp\gl}=\gl\frac{\pp<A>}{\pp\gl}$.  Thus the Legendre transform structure
of thermodynamics has been translated into the Fisher context (see here also
Chap. 4 of \cite{f4} for more on this and for additional general information we cite
e.g. \cite{ffpp,f1,f2,mppo,mnpp,mppv,masi,pepl,plpe,pppc,pppl,ppfi,plpi,pocu,ppnr,ppmk,plpo,
pcp,pppo,pppi,rcpo}).
\\[4mm]\indent
On the other hand
with $\bar{A}$ the sole constraint consider
\bq\label{5.15}
\tl{H}=FI +\ga(\bar{A}-\int pAdx)
\end{equation}
Then one finds directly
\bq\label{5.16}
\gd \tl{H}=\int dx\,\gd p\left[\frac{(p')^2}{p^2}+\pp_x\left(\frac{2p'}{p}\right)+\ga A\right]
\end{equation}
This means (as in (5.8))
\bq\label{5.17}
\frac{(p')^2}{p^2}+\pp_x\left(\frac{2p'}{p}\right)+\ga A=0
\end{equation}
Now we can write $\pp_x(2p'/p)=(2p''/p)-[2(p')^2/p^2]$ which means, via (4.5), that
\bq\label{5.18}
2\frac{\gD p^{1/2}}{p^{1/2}}+\ga A=0
\end{equation}
and via ({\bf 2B}) this means that the extreme probability $p_I$ directly determines a
quantum potential Q via
\bq\label{5.19}
Q=-\frac{\hbar^2}{2m}\frac{\gD p_I^{1/2}}{p_I^{1/2}}\Rightarrow Q=
\frac{\hbar^2}{4m}[\ga A]
\end{equation}
However, although A has not been specified we seem to have a result that a constraint $\bar{A}$ for which the
Fisher thermodynamic procedure works with $I[p]$ as defined, requires $A$ to
satisfy (5.19).  This is in fact tautological since we are dealing with a situation where fluctuations
based on $p_I$ are the only source of energy and one will have
$({\bf 5K})\,\,
\tl{F}=\int PQdx=(\hbar^2/8m)\int [(\na p)^2/p]dx$ (cf. \cite{c3}) and $\tl{F}$ corresponds to a 
fluctuation energy.
In particular as indicated in \cite{c3}
\bq\label{5.20}
\int p_IQdx=\frac{\hbar^2}{4m}\int dx\,p[\ga A]=\frac{\hbar^2}{4m}
[\ga\bar{A}]
\end{equation}
Note that in general
Fisher information $\tl{F}=I[p]$ as in ({\bf 5C})
is an action term which can be added to a classical Hamiltonian (as in Section 2) in order
to quantize it and thus $\tl{F}$ is a ``natural" constraint ingredient.  Fixing $\bar{A}$ would mean fixing the contribution of the quantum potential (QP) Q or fixing the amount
of quantization allowed.  In some way this would also 
correspond to restraining the probability 
in order to achieve
a fixed amount of quantization. 
\\[3mm]\indent
{\bf REMARK 5.1.}
Suppose we extremize $\tl{{\mf S}}$ as in (2.10) with $A\sim E=$ total energy to arrive at a
probability $({\bf 5L})\,\,P=(1/Z)exp(-\gag E)$ where $E\sim E(x,t),\,\,Z=\int exp(-\gag E)dx$.  Then compute
the Fisher information for this P which will be based upon (cf. (2.2))
\bq\label{5.21}
P'=\frac{1}{Z}(-\gag E')e^{-\gag E};\,\,P''=-\frac{\gag}{Z}[E''-\gag (E')^2]e^{-\gag E}
\end{equation}
\bq\label{5.22}
\frac{P'}{P}=-\gag E';\,\,\frac{P''}{P}=-\gag E''+\gag^2(E')^2
\end{equation}
\bq\label{5.23}
Q=\frac{\gag\hbar^2}{8m}[\gag(E')^2-2E'']
\end{equation}
\bq\label{5.24}
FI=\frac{8m}{\hbar^2}\int PQdx=\frac{\gag^2}{Z}\int (E')^2e^{-\gag E}dx
\end{equation}
On the other hand extremizing FI as in (5.15),
with fluctuations as the only source of energy, leads to $P=p_I$ defined via A as in (5.18) which
yields in paticular $Q=(\hbar^2/4m)[\ga A]$ as in (5.19).  Now recall for $P$ as in Section 3
\begin{enumerate}
\item
$\frac{P(x,t)}{P(x,0)}= e^{\frac{-\gD{\mc Q}}{kT}}$ from ({\bf 3C})
\item
$\gD{\mc Q}={\mc Q}(t)-{\mc Q}(0)=2\go[(\gd S)(t)-(\gd S)(0)]$ from (3.1) and $(2/\hbar)\gd S(t)
\sim \gb{\mc Q}(t)$ (cf. ({\bf 3T}))
\item
$\gd p=\na(\gd S(x,t))=-(\hbar/2)(\na P/P)$ as in (3.3) and from (3.13) $(\na P/P)=(1/\go\hbar)\na{\mc Q}=\gb\na {\mc Q}$
\item
$\gd E_{kin}=(\hbar^2/8m)(\na P/P)^2=(1/8m\go^2)(\na{\mc Q})^2$
\item
$P=\hat{c}(t)exp(-\ga{\mc Q})$ from (3.14) with $\ga=1/\go\hbar\sim 1/kT\sim\gb$ - this can also
be seen from (3.2) as $P\propto exp[-\gb{\mc Q}(t)]$
\end{enumerate}
For E = total energy as in ({\bf 5L}) it is clear that to apply this here we must think of $E=\hbar\go+\gd E_{kin}$ and $\hbar\go$ will only enter as a constant.  The thermalization fluctuation energy for 
probabilities appears as $\hat{c}exp[-\gd S(x,t)]\sim\hat{c}exp[-\gb{\mc Q}(t)]$ as indicated
so $\gd E_{kin}$ is expressed in terms of ${\mc Q}$.
This means that (5.24) involving $(E')^2$ (or $(\na E)^2$) for FI corresponds to
the $(\na{\mc Q})^2$ formula of (3.18).  In other words the fluctuation energy is equivalent to the
the thermal energy (modulo constants); see here also \cite{c3,gssn,hily}.$\hfill\bs$
\\[3mm]\indent
We omit here mention of many related topics and work involving Fisher information, thermodynamics, and entropy based on results of Abe, Frieden, Garbaczewski, Gellman,
Hall, Kaniadakis, Naudts, Pennini, A. Plastino, A.R. Plastino, Reginatto, Soffer, and Tsallis
in particular; some references can be found in \cite{c3,ebsv,f1,f2,f4,glts,ohpz}.

\end{document}